\documentclass[acmtog]{acmart}

\AtBeginDocument{%
  \providecommand\BibTeX{{%
    \normalfont B\kern-0.5em{\scshape i\kern-0.25em b}\kern-0.8em\TeX}}}

\setcopyright{acmcopyright}
\copyrightyear{2022}
\acmYear{2022}

\DeclareGraphicsExtensions{.pdf,.jpeg,.png}

\usepackage{hyperref}
\usepackage{todonotes}
\usepackage{stfloats}
\usepackage{bbding}

\begin{document}

\title{A Web-Based Tool for Comparative Process Mining}

\author{Madhavi Bangalore Shankara Narayana}
\authornote{Corresponding author.}
\email{madhavi.shankar@pads.rwth-aachen.de}
\orcid{0000-0002-7030-1701}
\affiliation{
  \institution{RWTH Aachen University}
  \streetaddress{Ahornstr. 55}
  \city{Aachen}
  \country{Germany}
  \postcode{52074}
}

\author{Elisabetta Benevento}
\authornotemark[1]
\email{benevento@pads.rwth-aachen.de}
\orcid{0000-0002-3999-8977}
\affiliation{
  \institution{RWTH Aachen University}
  \streetaddress{Ahornstr. 55}
  \city{Aachen}
  \country{Germany}
  \postcode{52074}
}
\affiliation{
  \institution{University of Pisa}
  \city{Pisa}
  \country{Italy}
}

\author{Marco Pegoraro}
\authornotemark[1]
\email{pegoraro@pads.rwth-aachen.de}
\orcid{0000-0002-8997-7517}
\author{Muhammad Abdullah}
\email{muhammad.abdullah1@rwth-aachen.de}
\author{Rahim Bin Shahid}
\email{rahim.shahid@rwth-aachen.de}
\author{Qasim Sajid}
\email{qasim.sajid@rwth-aachen.de}
\author{Muhammad Usman Mansoor}
\email{usman.mansoor@rwth-aachen.de}
\author{Wil M.P. van der Aalst}
\email{wvdaalst@pads.rwth-aachen.de}
\orcid{0000-0002-0955-6940}
\affiliation{
  \institution{RWTH Aachen University}
  \streetaddress{Ahornstr. 55}
  \city{Aachen}
  \country{Germany}
  \postcode{52074}
}

\renewcommand{\shortauthors}{M. B. Shankar N., E. Benevento, M. Pegoraro et al.}

\begin{abstract}
  Process mining techniques enable the analysis of a wide variety of processes using event data. Among the available process mining techniques, most consider a single process perspective at a time---in the shape of a model or log. In this paper, we have developed a tool that can compare and visualize the same process under different constraints, allowing to analyze multiple aspects of the process. We describe the architecture, structure and use of the tool, and we provide an open-source full implementation.
\end{abstract}



\maketitle

\section{Introduction}
Many companies today are unaware of why their processes are not performing at the maximum capacity, what are the bottlenecks in their processes, and which are the root causes of such performance issues. Process mining enables early detection of such problems, and allows companies to use it for monitoring and optimizing their process.

Comparing the behavior of a process under different circumstances, like time and other parameters, can help in identifying the causes of differences in models and performance. Particularly, comparative process mining aims at finding differences between two processes, and to detect process-related issues. Comparing the behaviour of two processes can help to investigate why one of them performs better in terms of predetermined characteristics or key performance indicators, to extract patterns or additional information, and to improve the understanding of these processes.

There exist two variants of process comparison: model-based or log-based comparison. The former uses models as input, while the latter considers event logs. Model-based comparison~\cite{cordes2014generic,la2013business,ivanov2015bpmndiffviz} is based on the control-flow dimension, and focuses on comparing the similarity of the input models. Model-based comparison techniques start by extracting a model from an event log through process discovery, and then check what activities are present in all models, or absent in one of them. Often, color coding or thickness of arcs and activities is used to highlight (significant) differences. The main drawback of model-based comparison is that the comparison is based mostly on the structure of the input models. It is not possible to analyze other process metrics (e.g., frequency or time statistics). Log-based comparisons do not have such limitation. For example, in~\cite{van2016log}, the proposed framework allows comparing the event logs of two processes. Such framework constructs a transition system based on the two-input event logs and provides process metrics. The coloring allows a user to quickly see where the differences of the two input event logs lie. In~\cite{van2013comparative}, the authors propose a comparative analysis using process cubes. Each cell in the cube corresponds to a set of events that can be used to discover a process model, to check conformance, or to discover bottlenecks. Slicing, rolling-up, and drilling-down, enable viewing event data from different angles and produce results that can be compared.

\begin{figure}[t]
\centering
\includegraphics[width=.7\linewidth]{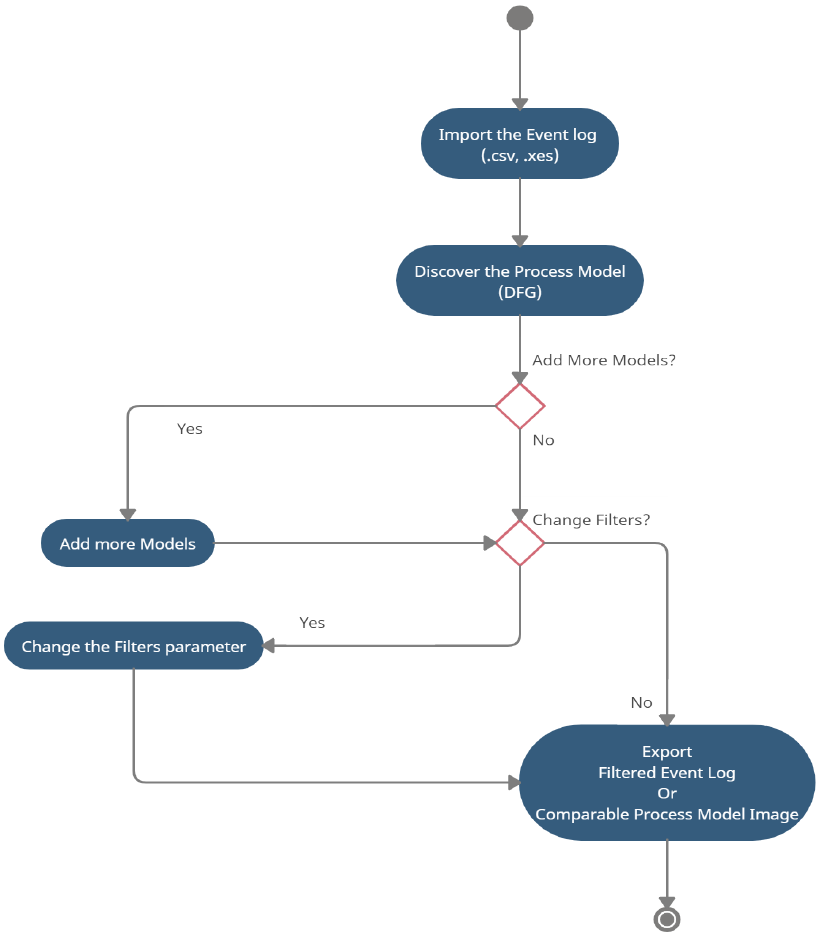}
\caption{Flowchart diagram of our tool.}
\label{fig:arch}
\end{figure}

\begin{figure*}[t]
\centering
\includegraphics[width=.67\textwidth]{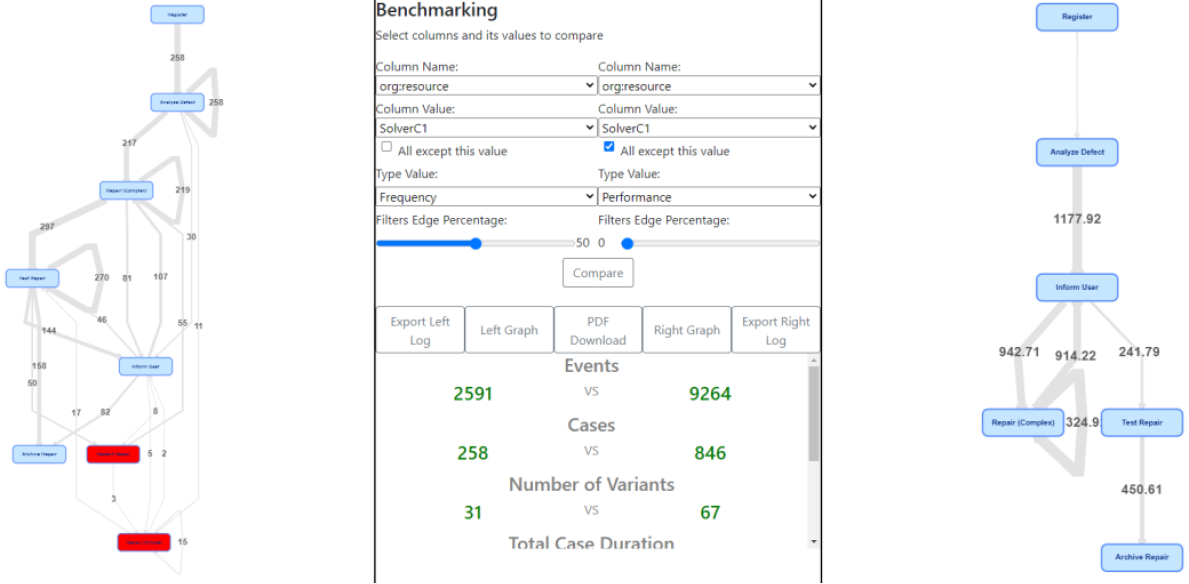}
\caption{Interface of the tool showing a comparison between two sections of an event log, obtained through filtering.}
\label{fig:cpm}
\end{figure*}

The purpose of our log-based comparison tool is to derive a way in which one can visualize the business process under certain conditions. Moreover, it also helps in comparing process models generated to find patterns and additional information to have a better understanding of the process.

The  remainder  of  this  paper  is  structured  as  follows. In Section~\ref{sec:description}, we describe the tool, its functions, and its architecture. 
Then, Section~\ref{sec:conclusion} concludes this paper.

\section{Architecture and Description}\label{sec:description}
The high-level architecture of the tool is shown in Figure~\ref{fig:arch}. The tool is build on Python 3.8 and Django framework. PM4Py 2.7.7.1 grants the core process mining functionalities, such as import/export of logs and models. We obtain Directly-Follows Graph (DFG) models and visualize them using the G6 library.

The tool allows the users to upload and/or select an event log from the root directory of event logs.
In the Comparison window, we provide users with the ability to choose a column and its attributes, which will be used to filter the log for variants satisfying the criteria. Users can choose what they wishes to visualize and compare. A common discovery algorithm will be applied to the variants, and once the models are generated, these will be analyzed to identify unique activities and edges. Unique elements in an activity will be highlighted so the user is easily able to identify comparative information. We also provide statistics such as number of traces, cases and average running time of the variants, and will be displayed to the user between the two filtered variants model. There, users can export the complete comparison information as PDF or can choose to export the individual components, like export variants, logs, or DFGs individually. The user may also add more models, although the comparison is limited to two models. The red colored activities in the model signify that these activities are not common between the models. The models can be visualized in performance or frequency metrics.
Figure \ref{fig:cpm} shows the comparison of a given process at two different time intervals. It also highlights the activities that are not common in both the models.

The implementation and user manual of the tool are available on Github\footnote{\url{https://github.com/MuhammadUsman05/Comparative-Process-Mining}}. A video is also available\footnote{\url{https://youtu.be/mNBGXM6mD14}}.
We have used a COVID-19 patients log recorded in the context of the COVID-19 Aachen Study (COVAS)~\cite{DBLP:conf/aktb/PegoraroSBAMM21} to demonstrate the tool. in this demonstration, we have compared the models in two different time ranges. Figure \ref{fig:cpm} shows the process model followed in Uniklinikum, Aachen for the first two waves of COVID-19.

\section{Conclusion and Future Work}\label{sec:conclusion}
In this paper, we presented our Comparative Process Mining tool built on the existing PM4Py framework. This tool will enable the research community to compare the behavior of processes in terms of performance or frequency and derive ideal and baseline models. As future work, we plan to enable comparing more than two models.

\begin{acks}
We acknowledge the ICU4COVID project (funded by European Union's Horizon 2020 under grant agreement n. 101016000) and the COVAS project for our research interactions.
\end{acks}

\bibliographystyle{ACM-Reference-Format}
\bibliography{bibliography}

\end{document}